\documentclass[prl,nofootinbib,superscriptaddress]{revtex4}
\usepackage[a4paper,left=1.5cm,right=1.5cm,top=3cm,bottom=3cm]{geometry}
\usepackage[toc]{appendix} %add option page to add a cover to the appendices

\usepackage{amssymb,amsmath,amsfonts}
\usepackage[dvipsnames]{xcolor}
\usepackage{graphicx}
\usepackage{longtable}
\usepackage{verbatim}
\usepackage{color}
\usepackage[mathscr]{euscript}
\usepackage{framed}
\usepackage{amsmath}
\usepackage{mdframed}
\usepackage{amsmath}
\usepackage{amsfonts}
\usepackage{mathrsfs}
\usepackage{amssymb}
\usepackage{mathrsfs}  
\newcommand{\arXiv}[2]{\href{http://arxiv.org/pdf/#1}{{\tt [#2/#1]}}}
\newcommand{\arXivold}[1]{\href{http://arxiv.org/pdf/#1}{{\tt [#1]}}}

\usepackage{color}
\usepackage{hyperref}
\hypersetup{colorlinks, citecolor=bluscuro, linkcolor=black, urlcolor=bluscuro}
\definecolor{rossos}{cmyk}{0,1,1,0.55}
\definecolor{bluscuro}{rgb}{0.15, 0.2, .85}
\definecolor{bluchiaro}{cmyk}{1,.3,0.,0.1}

\graphicspath{{./Figures/}}

\def\ltsim{\lower3pt\hbox{$\, \buildrel < \over \sim \, $}}  
\def\gtsim{\lower3pt\hbox{$\, \buildrel > \over \sim \, $}}  
\newcommand{\be}{\begin{equation}}  
\newcommand{\ee}{\end{equation}}  
\def\ga{\mathrel{\raise.3ex\hbox{$>$\kern-.75em\lower1ex\hbox{$\sim$}}}}  
\def\la{\mathrel{\raise.3ex\hbox{$<$\kern-.75em\lower1ex\hbox{$\sim$}}}}  

\makeatletter   
\@addtoreset{equation}{section}   
\makeatother   
   
\numberwithin{equation}{section}

\newcommand{\dd}{{\rm  d}}

\newcommand{\R}{\zeta}

\newcommand{\mail}[1]{\href{mailto:#1}{{#1}}}

\newcommand{\doi}[2]{\href{https://dx.doi.org/#1}{#2}}

\usepackage[dvipsnames]{xcolor}
\usepackage{hyperref}
\hypersetup{colorlinks, citecolor=ForestGreen, linkcolor=BrickRed, urlcolor=NavyBlue}
\makeatletter   
\@addtoreset{equation}{section}   
\makeatother

\usepackage[normalem]{ulem}

\begin{document}  
  
\baselineskip=18pt   
\begin{titlepage}  
%\rightline{}  
\begin{center}  
\vspace{0.5cm}  
  
\Large {\bf The Primordial Black Hole Formation \\ from Single-Field Inflation
is Still Not Ruled Out}
  
\vspace*{20mm}  
\normalsize  

{\large\bf 
 Antonio ~Riotto\footnote{\mail{antonio.riotto@unige.ch}}  }

\smallskip   
\medskip

\it{\small $^1$ Department of Theoretical Physics and Gravitational Wave Science Center,  \\
			24 quai E. Ansermet, CH-1211 Geneva 4, Switzerland}

%\smallskip   
  
  \end{center}
  
\vskip0.6in

\centerline{\bf Abstract}
\vskip 0.5cm  
\noindent
In response to a recent criticism, appeared in Ref. \cite{yokoyamanew}, we argue that the 
 standard scenario to form primordial black holes in the early universe  based on a  phase of ultra-slow-roll in single-field inflation is not ruled out.

\vspace*{2mm}   
%\smallskip\newline  

\end{titlepage} 

%%%%%%%%%%%%%%%%%%%%%%%%%%%%%%%%%%%%%%%%%%%%%%%%%%%%%%%%%%%%
%%%%%%%%%%%%%%%%%%%%%%%%%%%%%%%%%%%%%%%%%%%%%%%%%%%%%%%%%%%%
\section{I. Introduction and Conclusion}  \label{sec:intro}
\setcounter{section}{1}
\noindent
 A  standard  mechanism to create   Primordial Black Holes (PBHs) in the early universe  is  by enhancing   the short scale curvature perturbation $\zeta_S$   \cite{s1,s2,s3} during an Ultra Slow Roll (USR)  period when  the inflaton potential  $V(\phi)$ is quite flat. 
If loop  corrections from the short modes are larger than the tree-level  contribution to the large-scale power spectrum $P_\zeta(k_L)$, the USR mechanism to generate PBHs will be inevitably  ruled out,  as  advocated in Ref. \cite{yokoyama}, and more recently in Ref. \cite{yokoyamanew}. In a short note, we have  argued 
that the PBH Formation  from single-field inflation
is in fact not ruled out \cite{riotto}. The argument was as follows.  Short modes propagate in a universe perturbed by the presence of a long, basically static,  CMB mode which can be absorbed by a short momentum rescaling. The loop correction  is therefore dictated by the correlation between the short mode expectation value and the long wavelength mode or, in other words,  by the corresponding squeezed bispectrum

\be
\label{fund1}
\langle \R_L\R_L\rangle\sim \langle\langle\zeta_S\zeta_S\rangle_{\zeta_L}\zeta_L \rangle\sim  \frac{\dd\ln {\cal P}_\zeta(k_S)}{\dd\ln k_S}P_\zeta(k_S)P_\zeta(k_L),
\ee
Loop corrections are  therefore  suppressed by the breaking of scale invariance in the short mode power spectrum or, in other words, by field derivatives of the inflaton potential. In Ref. \cite{yokoyamanew} this point has been questioned based on the fact that a non-suppressed (by slow-roll parameters) squeezed bispectrum  (corresponding to a non-linear non-Gaussianity parameter $f_{\rm NL}=5/2$) is present once one recall that the cubic interaction 

\be
\label{a}
S_{\rm int}[\R]\sim \int\dd^3x\,\dd t\,a^3\epsilon\,\frac{1}{2}\dot \eta \dot \R\R^2
\ee
is obtained after the field redefinition $\zeta\rightarrow\zeta+\eta\,\zeta^2/4+\zeta	\zeta'$ necessary to remove a boundary term    \cite{m}.

However, in Ref. \cite{riotto} (see end of Section III) it was  
explicitly  stated that this is not necessarily  true when taking into account the details  of the transition into the slow-roll phase following the USR period. Let us elaborate about this point further here. If the transition is not sudden, the non-Gaussianity evolves to become suppressed by slow-roll parameters and the argument of Ref. \cite{riotto} applies. 
To show this explicitly,  we follow Ref. \cite{c} and write the interaction term (\ref{a}) in the form

\be
\label{b}
S_{\rm int}[\R]\sim -\int\dd^3x\,\dd t\,\frac{\dd}{\dd t}\left(\frac{a^3\epsilon\dot \eta}{6}\right)\,\R^3.
\ee
Suppose  that the transition is not sudden, in the sense that the slow-roll parameter $\epsilon_V$ during the slow-roll phase might be  smaller compared to the inflaton velocity at the end of the USR phase. This can be achieved for instance in the Starobinsky spiky model  \cite{sta} where there is an abrupt change in the slop of the linear potential of the inflaton field $\phi$. Even if one is subjects to  drastic variations of the $\eta$ parameter during the USR phase, inserting the  analytical  solutions for $\epsilon$ and $\eta$ one finds \cite{c}

\be
-\frac{\dd}{\dd t}\left(\frac{a^3\epsilon\dot \eta}{6}\right)=\frac{a^3\epsilon}{3}\sqrt{2\epsilon}V'''.
\ee
Therefore, the  contribution from this term is always tiny, no matter how large is  $\eta$  and $\dot \eta$ are during the transition, and in fact suppressed by slow-roll parameters. Indeed, the corresponding squeezed bispectrum $\langle\langle\zeta_S\zeta_S\rangle_{\zeta_L}\zeta_L \rangle$ is suppressed by ${\cal O}(\epsilon_V^{1/2}\eta_V)$ \cite{c}, with $\epsilon_V$  small enough to  generate PBHs.
In other words, if  the curvature perturbations keep evolving during the transition from the USR phase into the slow-roll phase (this happens in realistic cases of    smooth transitions or some sharp transition scenarios), the  large local non-Gaussianity generated in the USR phase is erased by the subsequent evolution and the argument of Ref. \cite{riotto} applies.

As a closing note, let us also add an important point that in the discussion has never been touched upon. Much ambiguity in the literature has been  present regarding  which  perturbations one should use to compute the   PBH abundance. However, it is rather clear by now that the most correct correct quantity is  the smoothed density contrast 
\cite{Y}
\be
\delta_m=\delta_l-\frac{3}{8}\delta_l^2,\,\,\,\,
\delta_l=-\frac{4}{3}r_m\zeta'(r_m),
\ee
where $r_m$ is the point where the compaction function has its maximum, 
rather than the  metric perturbation $\zeta$. Indeed, on  superhorizon scales  one can always shift the comoving curvature perturbation by an arbitrary constant by a coordinate transformation. This  makes the calculation of the PBH unphysical. This problem is avoided by using   $\delta_m$  which depends on spatial derivatives of the curvature perturbation.

The point is now that the PBH abundance calculated through the correct variable $\delta_m$ requires much smaller values of the short scale power spectrum  amplitude ${\cal P}_\zeta(k_S)$ than if the same PBH abundance is calculated using the comoving curvature perturbation. For instance, for the value adopted in Ref. \cite{yokoyamanew} ${\cal P}_\zeta(k_S)={\cal O}(10^{-1})$, the same abundance of PBHs, when calculated with the correct variable $\delta_m$, requires ${\cal P}_\zeta(k_S)={\cal O}(10^{-3})$ \cite{DR}, which reduces considerably the loop correction to the large scale CMB spectrum even in the case of sudden transition.

We conclude that the one may not  claim that the PBH formation scenario  from single-field inflation
is  ruled out in full generality as the fine details   of the transition from the USR phase into the standard slow-roll one matter, as well as the adoption of the correct variable
to calculate the PBH abundance.

\vskip 0.5cm
\centerline{\bf Acknowledgements}
\vskip 0.2cm
\noindent
A.R. is supported by the
Boninchi Foundation for the project ``PBHs in the Era of
GW Astronomy".

%%%%%%%%%%%%%%%%%%%%%%%%%%%%%%%%%%%%%%%%%%%%%%%%%%%%%%%%%%
%%%%%%%%%%%%%% REFERENCES %%%%%%%%%%%%%%%%%%%%%%%%%%%%%%%%

\end{document}